%% file: main.tex
\documentstyle[12pt,psfig]{article}
\def\rmo{r_{\rm o}}
\def\rmot{r^3_{\rm o}}
\def\1#1{{\bf #1}}
\def\lp{\left(}
\def\rp{\right)}

\begin{document}

\begin{center}

\vspace{1.0cm}

{\large \bf Van der Waals Excluded Volume Model for\\
Lorentz Contracted Rigid Spheres
}

\vspace{1.0cm}

{\bf K.A. Bugaev$^{1,2}$, M.I. Gorenstein$^{1,2}$,
H. St\"ocker$^{1}$ and W. Greiner$^{1}$
}
\end{center}

\vspace{1.cm}
\noindent
$^1$Institut f\"ur Theoretische Physik,
Universit\"at Frankfurt, Germany

\vspace{0.2cm}
\noindent
$^2$Bogolyubov Institute for Theoretical Physics,
 Kiev, Ukraine

\vspace{2.0cm}
\begin{center}
{\bf Abstract}
\end{center}
Conventional cluster and virial expansions
are generalized to momentum dependent
inter-particle potentials.
The model with 
Lorentz contracted hard core 
potentials
is considered, e.g. as hadron gas model.
A Van der Waals-type model
with a temperature dependent excluded volume is derived.
Lorentz contraction effects at given temperature
are stronger for light particles and make their effective
excluded volume smaller than that of heavy
ones.

\vspace{4.cm}


\newpage

The Van der Waals (VdW) excluded volume model has been used
to describe hadron yields in relativistic nucleus--nucleus
collisions (see e.g. \cite{Yen:99,St:99} and references therein). 
This model treats the hadrons as hard-core spheres and, therefore,
takes into account the hadron
repulsion at short distances. In a relativistic situation one should,
however,
include the Lorentz contraction of the hard core-hadrons.
This problem was discussed in the literature (see e.g. 
Ref.~\cite{Kap:83,Z:95}).
In this paper the cluster and virial expansions are generalized to
 velocity dependent inter-particle potentials. This extension is
used to construct the VdW model for Lorentz contracted
rigid spheres which may be used to simulate
hadrons.
 
The canonical partition function for the gas of $N$ classical (Boltzmann)
particles takes the form 
\begin{equation}\label{cpf}
Z_N(V,T)~=~\frac{1}{N!}~\int\prod_{i=1}^{N}\left[\frac{g~d{\bf r}_i
d{\bf k}_i}{(2\pi)^3}~\exp\left(- \frac{\omega_i}{T}\right)\right]~
\exp\left(- \frac{U}{T}\right)~,
\end{equation}
where $V$ and $T$ are the system volume and temperature,
$g$ is the number of internal degrees of freedom 
(degeneracy factor) of the particles,
$\omega_i=(m^2+{\bf k}_i^2)^{1/2}$ is the dispersion
relation of free particles
with masses $m$. The particle interactions described by the function
$U$ in Eq.~(\ref{cpf}) are given by
the sum
over pair potentials:
\begin{equation}\label{U}
U~=~\sum_{1\le i <j \le N}~u_{ij}~.
\end{equation}
In contrast to the usual statistical mechanic treatment of
the pair potentials, the $u_{ij}$ are assumed to be both
coordinate and momentum dependent 
$u_{ij}\equiv u({\bf r}_{i},{\bf k}_i;~{\bf r}_j,{\bf k}_j)$.
This generalization is necessary, if Lorentz 
contraction effects of hard spheres are to be taken into account. 
Introducing the Mayer functions 
\begin{equation}\label{fij}
f_{ij}~=~
\left[\exp \left(- \frac{u_{ij}}{T}\right) ~-~1\right]~,
\end{equation}
Eq.~(\ref{cpf}) can be presented as
\begin{equation}\label{cpf1}
Z_N(V,T)~=~\frac{1}{N!}~\int d{\bf x}_1 ... d{\bf x}_N~
\exp\left(-\frac{\omega_1+...+\omega_N}{T}\right)~
\prod_{1\le i < j \le N} (1~+~f_{ij})~,
\end{equation}
with the short notation
$
d{\bf x}_i\equiv g d{\bf r}_i d{\bf k}_i/(2\pi)^3.
$
Similarly to the standard procedure one can introduce
the cluster integrals \cite{May:77} 
\begin{eqnarray}
b_1~&=&~\frac{1}{V}~\int d{\bf x}_1 ~ \exp\left(-\frac{\omega_1}{T}
\right)~=~\frac{g~T^3}{2\pi^2}~K_2\left(\frac{m}{T}\right)~
\equiv ~\phi(T)~,
\label{b1}\\
b_2~&=&~\frac{1}{2!V}~\int d{\bf x}_1 d{\bf x}_2~
\exp\left(-\frac{\omega_1 +\omega_2}{T}\right)~f_{12}~,
\label{b2}\\
b_3~&=&~\frac{1}{3!V}~\int d{\bf x}_1d{\bf x}_2d{\bf x}_3~
\exp\left(-\frac{\omega_1+\omega_2+\omega_3}{T}\right)~
(f_{12}f_{13}~ \label{b3}\\
&+&~f_{12}f_{23}~+~f_{13}f_{23}~+~
f_{12}f_{23}f_{13} 
)~,\nonumber \\
... & & \nonumber
\end{eqnarray}
and present the canonical partition function 
in the familiar form
\begin{equation}\label{cpf2}
Z_N(V,T)~=~\sum_{\{m_l\}} ^{~~~~\prime}~
\prod_{l=1}^N ~\frac{(Vb_l)^{m_l}}{m_l!}~,
\end{equation}
where the summation  in Eq.~(\ref{cpf2}) is taken
over all sets of non-negative integer numbers $\{m_l\}$ 
satisfying the condition
\begin{equation}\label{cond}
\sum_{l=1}^N lm_l~=~N~.
\end{equation}
Note, however, that the cluster integrals defined above are
different from those used in standard statistical mechanics \cite{May:77}
as here nontrivial momentum integrations are included.  
Condition (\ref{cond}) makes the calculation of $Z_N$ (\ref{cpf2})
rather complicated. This problem can be avoided in the grand
canonical ensemble:
the grand canonical partition function can be calculated
explicitly ($z\equiv
\exp(\mu/T)$):
\begin{equation}\label{gcpf}
{\cal Z}(V,T,\mu)~\equiv ~\sum_{N=0}^{\infty}
\exp\left(\frac{\mu N}{T}\right)~
Z_N(V,T)~=~\exp\left(~V~\sum_{l=1}^{\infty}b_l z^l~\right)~.
\end{equation}
In the thermodynamical limit the pressure $p$ and particle number density
$n$
are calculated in the grand canonical ensemble
in terms of the asymptotic values of the cluster integrals:
\begin{eqnarray}
p~&=&~T~\lim_{V \rightarrow \infty}~\frac{\ln {\cal
Z}}{V}~=~T~\sum_{l=1}^{\infty}b_l z^l~,
\label{pgce}\\
n~&=&~\lim_{V \rightarrow \infty}~\frac{1}{V}~
\frac{\partial \ln {\cal Z}}{\partial z}~=~
\sum_{l=1}^{\infty}l b_l z^l~. \label{ngce}
\end{eqnarray}
The virial expansion represents the pressure in terms of a series
of particle number density and takes the
form
\begin{equation}\label{virial}
p~=~T~\sum_{l=1}^{\infty}a_ln^l~.
\end{equation}
Substituting $p$ (\ref{pgce}) and $n$ (\ref{ngce})
into Eq.~(\ref{virial}) and equating the coefficients
of each power of $z$, one finds the virial coefficients $a_l$ in terms of
the cluster integrals 
\begin{equation}\label{al}
a_1~=~1~,~~~a_2~=~- \frac{b_2}{b_1^2}~,~~~a_3~=~\frac{4b_2^2}{b_1^4}~-
~\frac{2b_3}{b_1^3}~,~~~...
\end{equation}

\vspace{0.2cm}
Let us recall, first, the derivation of the standard 
VdW excluded volume model.
Then it is extended by adding the Lorentz contraction of the moving
particles. Keeping the first two terms of the virial expansion
(\ref{virial}) the following result is obtained:
\begin{equation}\label{second}
p(T,n)~=~Tn~(1~+~a_2n)~.
\end{equation}
It is valid for small particle densities (i.e. $n<<1/a_2$).
The usual (momentum independent) hard core potential 
for spherical particles with radius $\rmo$ is
$u_{ij}=u(|{\bf r}_i - {\bf r}_j|)$. Here the function $u(r)$
equals to 0 for $r>2 \rmo $ and $\infty$ for $r<2 \rmo$.
The second cluster integral (\ref{b2}) can easily be calculated
in this case:
\begin{equation}\label{b2h}
b_2~=~-~\phi^2(T)~\frac{16\pi}{3}\rmot~.
\end{equation}
Therefore $a_2=4\,v_{\rm o}$, where $v_{\rm o}=4\pi \rmot/3$
is the particle hard core volume.
The VdW excluded volume model is obtained as 
the extrapolation of Eq.~(\ref{second}) to large particle
densities in the form
\begin{equation}\label{vdw}
p(T,n)~=~\frac{Tn}{1~-~a_2n}~.
\end{equation}
For practical use the pressure is given
as a function of $T$ and $\mu$ independent
variables, i.e. in the grand canonical ensemble.
This is done by substituting $n=(\partial p/\partial \mu)_T$
into Eq.~(\ref{vdw}), which then turns into a partial differential
equation for the function $p(T,\mu)$.  For the VdW
model
(\ref{vdw}) the solution of this partial differential equation
can be presented in the form of a transcendental equation
\begin{equation}\label{vdwgc}
p(T,\mu)~=~T\phi(T)e^{\mu/T}\exp\left(-\frac{a_2p}{T}\right)~\equiv
~p_{id}(T,\mu-a_2p)~.
\end{equation}
Eq.~(\ref{vdwgc}) was first obtained in Ref. \cite{Dirk:91}
using the Laplas transform technique.
With $p(T,\mu)$ (the solution of Eq.~(\ref{vdwgc}))
the particle number density, entropy density and energy density
are calculated as ($\nu =\mu-a_2\,p(T,\mu),\,\, a_2=4\,v_{\rm o}$):
\begin{eqnarray}
n(T,\mu)~&\equiv &~\left(\frac{\partial p(T,\mu)}{\partial
\mu}\right)_{T}~=
~\frac{n_{id}(T,\nu)}{1~+~a_2n_{id}(T,\nu)}~,\label{ntmu}\\
s(T,\mu)~&\equiv &~\left(\frac{\partial p(T,\mu)}{\partial
T}\right)_{\mu}~=
~\frac{s_{id}(T,\nu)}{1~+~a_2n_{id}(T,\nu)}~,\label{stmu}\\
\epsilon(T,\mu)~&\equiv &~Ts~-~p~+~\mu n~=~
\frac{\epsilon_{id}(T,\nu)}{1~+~a_2n_{id}(T,\nu)}~.\label{etmu}
\end{eqnarray}
Here the superscripts $id$ in the thermodynamical functions
(\ref{vdwgc}--\ref{etmu}) indicate those of the ideal gas.

The excluded volume effect
accounts for the blocked volume 
of two spheres when they touch each other.
If hard-sphere particles move with relativistic velocities
it is necessary to include their Lorentz contraction
in the rest frame of the fluid.
The model suggested in Ref. \cite{Z:95} 
is not satisfactory: the parameter 
$a_2=4\,v_{\rm o}$ of the VdW excluded volume model
is confused there
with the proper volume of an individual particle --
the contraction effect is introduced for 
the proper volume of each particle. 
In order to get the correct result it is necessary to 
account for the excluded volume of
two Lorentz contracted spheres.

Let ${\bf r}_i$ and ${\bf r}_j$ 
be the coordinates of the $i$-th and $j$-th particle, respectively, 
and ${\bf k}_i$ and ${\bf k}_j$ be their momenta,
${\bf {\hat r}}_{ij}$ denotes 
the unit vector
$ {\bf {\hat r}}_{ij} = {\bf r}_{ij}/|{\bf r}_{ij}|$
($ {\bf r}_{ij}= |{\bf r}_i -  {\bf r}_j|$).
Then for a given  
set of vectors $\left( {\1 {\hat r}}_{ij} , \1 k_i, \1 k_j \right)$
for the Lorentz contracted rigid spheres of radius $\rmo$
there exists the minimum distance between their centers 
$r_{ij} ({\bf {\hat r}}_{ij}; {\bf k}_i, {\bf k}_j) = {\rm min}|\1 r_{ij}|$.
The dependence of the  potentials 
$u_{ij}$ on the coordinates ${\bf r}_i,{\bf r}_j$
and momenta ${\bf k}_i, {\bf k}_j$ 
can be given in terms of the minimal distance
as follows
\begin{equation}
u({\bf r}_{i},{\bf k}_i; {\bf r}_j,{\bf k}_j)  =
\left\{ \begin{array}{rr}
0\,,  &\hspace*{0.3cm}|\1 r_i - \1 r_j| >   r_{ij} \lp
{\1 {\hat r}}_{ij}; \1 k_i, \1 k_j
 \rp  \,, \\
 & \\
\infty\,,  &\hspace*{0.3cm}|\1 r_i - \1 r_j| \le  r_{ij} \lp
{\1 {\hat r}}_{ij}; \1 k_i, \1 k_j
 \rp  \,.
\end{array} \right. 
\end{equation}

The general approach to the cluster- and virial expansions described above
is valid for this momentum dependent potential, and
it leads to Eqs.~(\ref{vdw},\ref{vdwgc}) with
\begin{eqnarray}
a_2(T)~&=&~\frac{1}{2\phi^2(T)}~\int \frac{d{\bf k}_1d{\bf k}_2~}
{(2\pi)^6}~
\exp\left(-\frac{\omega_1~+~\omega_2}{T}\right)~\times \label{a2rel}\\
&\times& ~\int d{\bf r}_{12}~\Theta\left(r_{12}
({\bf {\hat r}}_{12}; {\bf k}_1, {\bf k}_2)~-
~|{\bf r}_{12}|\right)~.\nonumber
\end{eqnarray}
The new feature is the temperature dependence of the excluded
volume $a_2$ (\ref{a2rel}) which is due to the Lorentz contraction
of the rigid spheres. The pressure and
particle number density are still given by Eqs.~(\ref{vdwgc},\ref{ntmu}), 
but with temperature dependent $a_2(T)$ (\ref{a2rel}). However,
Eqs.~(\ref{stmu},
\ref{etmu}) are now modified, e.g.
\begin{equation}\label{etmurel}
\epsilon(T,\mu)~=~\frac{\epsilon_{id}(T,\nu)
~-~p^2~da_2(T)/d T}
{1~+~a_2n_{id}(T,\nu)}~.
\end{equation}
In contrast to Eq.~(\ref{etmu}) the energy density (\ref{etmurel})
contains the extra term which appears also in the entropy density.
The excluded volume $a_2(T)$ (\ref{a2rel}) is always smaller than
$4\,v_{\rm o}$. It has been proven rigorously that $a_2(T)$
is a monotonously decreasing function of $T$
and, therefore, the additional term in Eq.~(\ref{etmurel}) is always
positive. 
Let us introduce the notation
\begin{equation}\label{f}
a_2(T)~=~4\,v_{\rm o}~f(T)~.
\end{equation}
The function $f(T)$ depends on the $T/m$ ratio. It can be calculated
numerically and its behavior
is shown in Fig.~1.
The simple analytical formula
\begin{equation}\label{f1}
f(T)~=~c~+~(1-c)~\frac{\rho_s(T)}{\phi(T)}
\end{equation}
with
$$ 
c~=~\left(1~+~\frac{74}{9\pi}\right)^{-1}~,~~~~
\rho_s~=~\frac{g}{(2\pi)^3}\int d{\bf k}~\frac{m}{\omega}~
exp\left(- \frac{\omega}{T}\right)~,
$$
is found to be valid with an accuracy of
a few percents for all temperatures.
The asymptotic behavior of $f(T)$ is the following:
 $1-{\rm O}(T/m)$ at
$T<<m$
and $c+{\rm O}(m/T)$ at $T>>m$.

If one assumes 
that all types
of hadrons 
have at rest the same hard core radius then 
the Lorentz contraction effect leads to different
VdW excluded volumes for moving particles with different masses:
for light particles (e.g. pions) the excluded volume
(at given $T$)  is smaller than
that for heavy ones. 
Fig.~1 shows that at $T\cong 150$~MeV the value
of $a_2$ in the nucleon gas ($m\cong 939$~MeV) decreases by 10\% in
comparison to its nonrelativistic
value $4\,v_{\rm o}$, whereas for pions ($m\cong 140$ MeV) 
$a_2$ shrinks at the same $T$ by almost a factor 2.
This is simply because light particles are more relativistic
than heavy ones at given temperature typical for high
energy nuclear collisions, $T=120\div 170$~MeV.

As an example, Fig.~2 shows  the particle number density
of the pion gas ($\mu=0, g=3$) with $r_{\rm o}=0.5$~fm.
The particle number density is calculated according to Eq.~(\ref{ntmu})
for three different models: the ideal pion gas ($a_2=0$),
the VdW model with constant excluded volume ($a_2=4v_{\rm o}$)
and the VdW model with
Lorentz contraction ($a_2(T)$ is given by Eq.~(\ref{a2rel})).
It can be seen from Fig.~2
that at low $T$ the pion density is small
and excluded volume corrections are unimportant. Therefore, 
all three models are similar. The situation
changes with increasing $T$: the suppression due to
the excluded volume effects are large and different
for $a_2=4v_{\rm o}$ and $a_2(T)$ (\ref{a2rel}).
The ratios of particle number densities and energy densities
of the pion gas
for two versions of the 
VdW model ($a_2=4v_{\rm o}$ and $a_2(T)$ (\ref{a2rel}))
are shown in Fig.~3 as functions of the temperature. From 
Fig.~3 one can observe the deviations between these two
models. These deviations increase with temperature. They are
larger for the energy density due to the additional positive term
in Eq.~(\ref{etmurel}). 

In conclusion, the traditional cluster and virial expansions
can be consistently generalized to momentum dependent pair potentials.
Hard-core potentials with Lorentz contraction effects
lead to a VdW model
with a temperature dependent excluded volume $a_2(T)$ (\ref{a2rel}).
For light particles the effect of Lorentz contraction
is, evidently, stronger than for heavy ones. Note that smaller
values of the pion hard core radius $r_{\pi}$ were introduced in
Refs.~\cite{Rit:97, Yen:97} within the standard VdW excluded volume model
to fit hadron yield data better. 
The smaller value of the pion excluded volume appears
as a consequence of stronger Lorentz contraction
for light particles.


\vspace*{1cm}

\begin{center}
{ \bf Acknowledgments}
\end{center}
\vspace{0.2cm}

The authors would like to thank 
R. Pisarski
for valuable comments and L. McLerran,  D.H. Rischke
and R. Venugopalan for discussions.
K.A.B. gratefully acknowledges
the warm hospitality of the BNL Nuclear Theory Group
and the BNL-RIKEN Center,
where parts of this work were done.
M.I.G. acknowledges financial support of DFG, Germany.
K.A.B. is grateful to the Alexander von Humboldt Foundation for   
the financial support.



\input vdwplots.tex

\end{document}

%% file: vdwplots.tex
\clearpage
\newpage

\begin{figure}

\vspace*{-1.0cm}

\mbox{
\hspace*{3.0cm}\psfig{figure=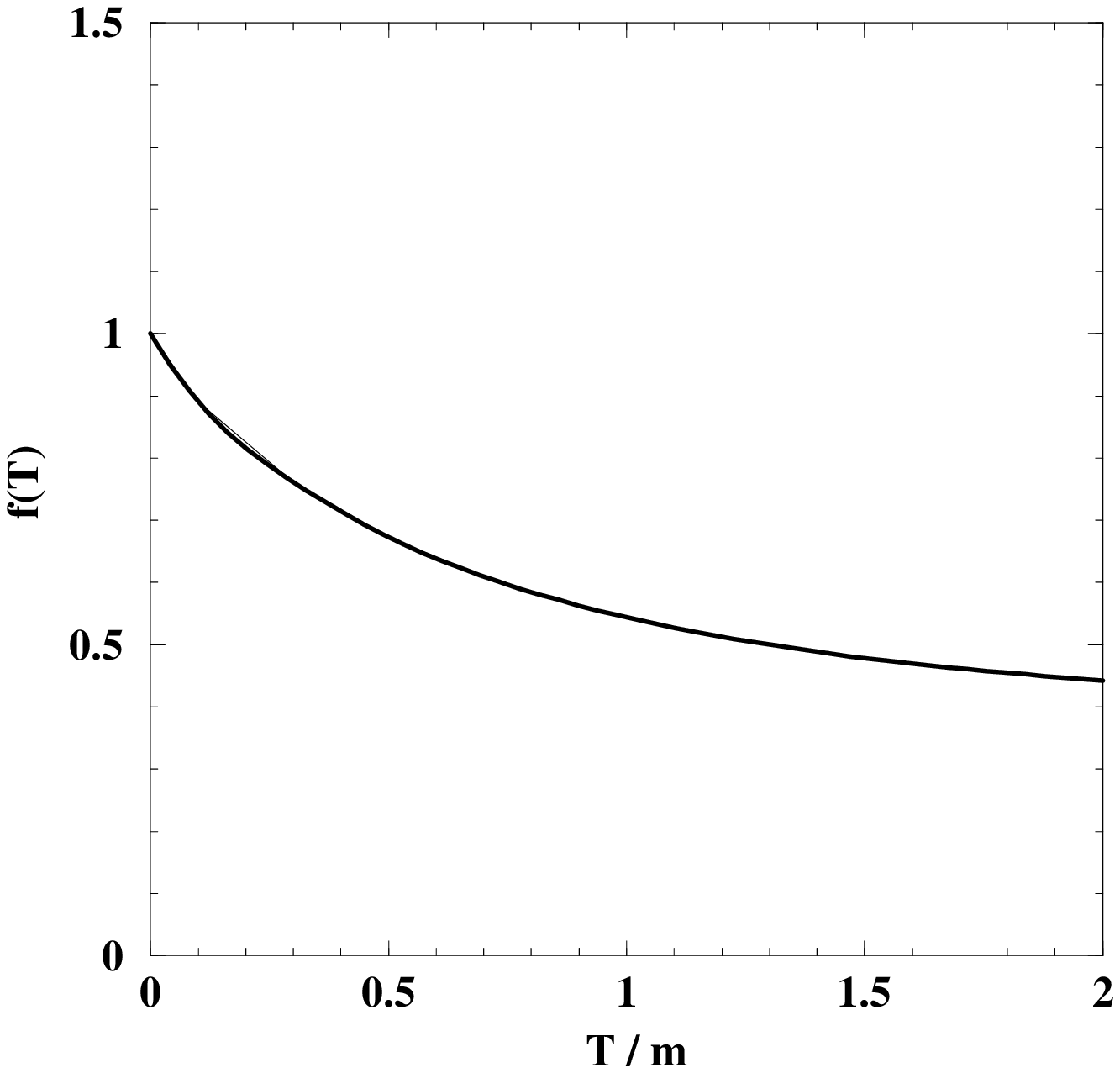,height=9cm,width=9cm}
}

\vspace*{-1.0cm}

{\bf Fig. 1.}
$f(T)$ 
as the function of the temperature-to-mass ratio. 
For heavy particles (e.g., nucleons $m >> T$) the volume reduction is just a few per cents,
whereas for pions ($m \approx T$) it is about 50\%. 

\end{figure}


\clearpage
\newpage
\begin{figure}

\vspace*{-1.0cm}

\mbox{
\hspace*{3.0cm}\psfig{figure=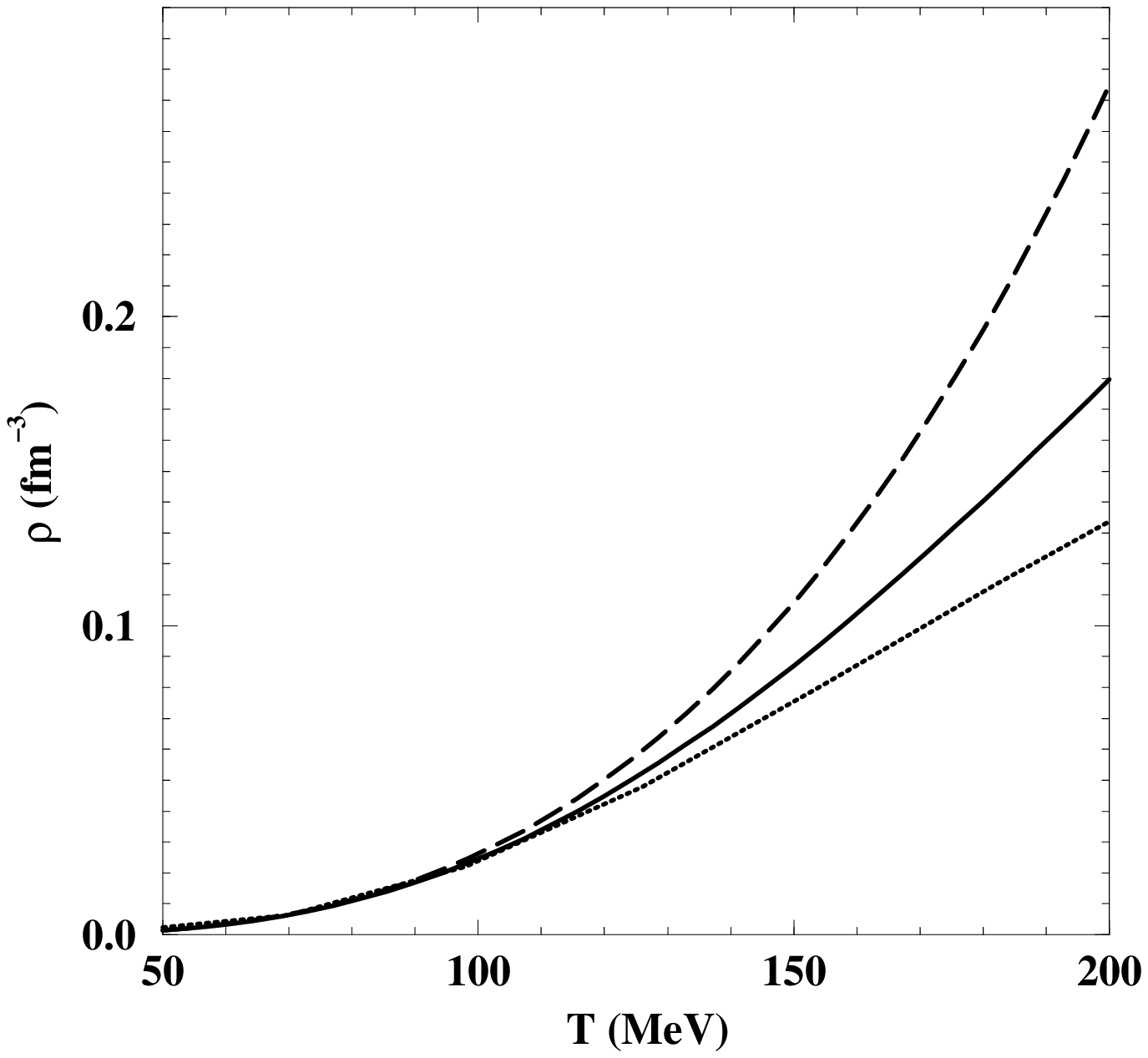,height=9cm,width=9cm}
}

\vspace*{-1.0cm}

{\bf Fig. 2.}
The  particle number  density
for three models of the pion gas ($\mu =0, g=3$): the
solid line corresponds to the VdW model of the Lorentz 
contracted spheres ($r_{\rm o}=0.5$~fm),
the dashed one  corresponds to the ideal gas of
point-like particles,
and the dotted one corresponds to
the VdW model without Lorentz contraction for the spheres of a constant
radius 0.5~fm.
\end{figure}


\clearpage
\newpage
\begin{figure}

\vspace*{-1.0cm}

\noindent
\mbox{
\hspace*{3.0cm}\psfig{figure=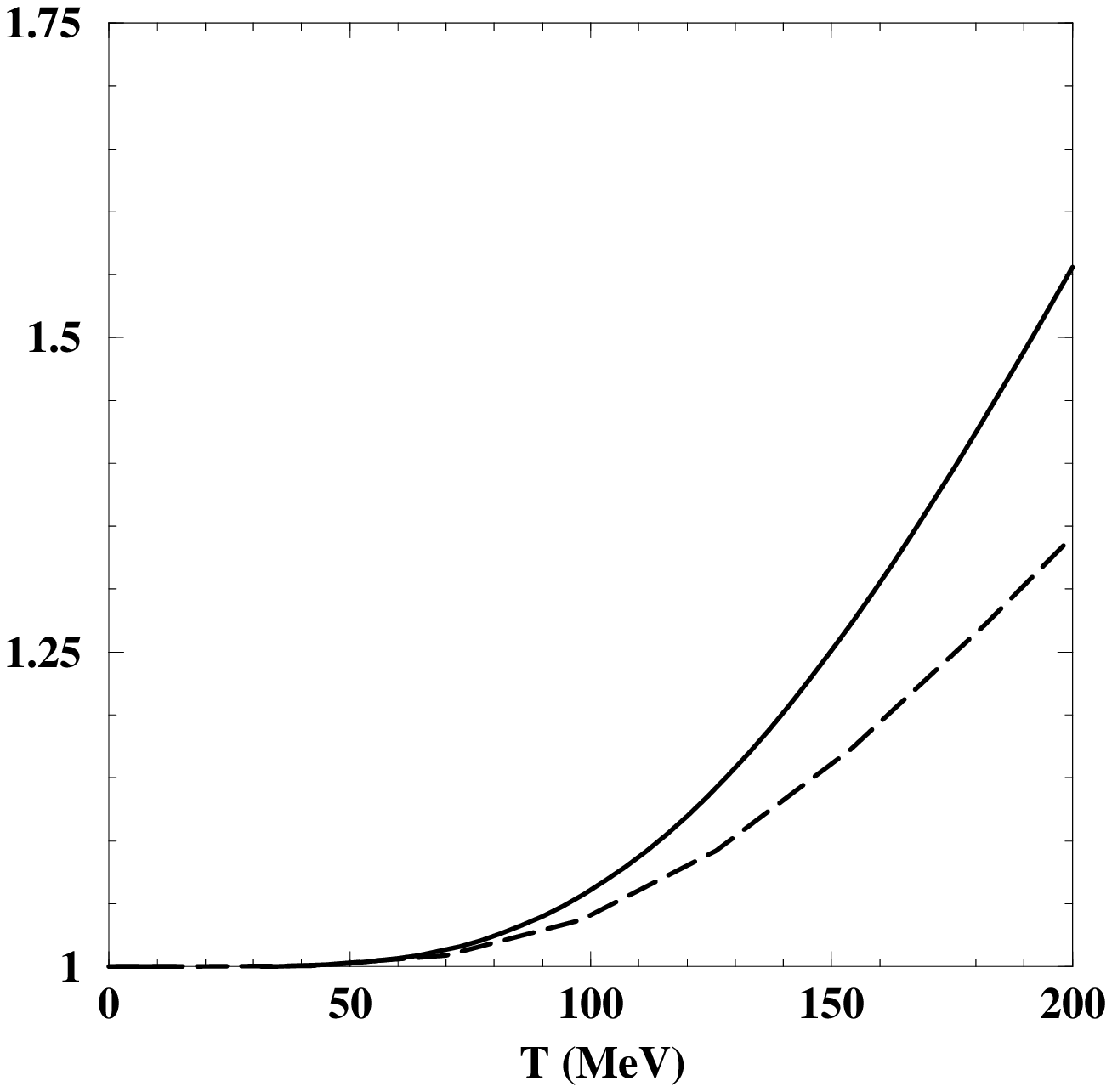,height=9cm,width=9cm}
}

\vspace*{-1.0cm}

{\bf Fig. 3.}
The dashed line shows the ratio of the particle number densities
of the pion gas
($g=3, \mu=0, r_{\rm o}=0.5$~fm): the VdW model with Lorentz
contraction divided by  the VdW model without Lorentz contraction.
The solid line shows a similar ratio for the energy densities.

\end{figure}

